\documentclass[showpacs,prl,twocolumn,groupedaddress]{revtex4}
\usepackage{graphicx}
\usepackage{dcolumn}
\usepackage{bm}
\usepackage{natbib}
\usepackage{amsmath}

\begin{document}

\title{Photon solid phases in driven arrays of nonlinearly coupled cavities}

\author{Jiasen Jin}
\affiliation{NEST, Scuola Normale Superiore and Istituto di Nanoscienze-CNR, I-56126 Pisa, Italy}
\author{Davide Rossini}
\affiliation{NEST, Scuola Normale Superiore and Istituto di Nanoscienze-CNR, I-56126 Pisa, Italy}
\author{Rosario Fazio}
\affiliation{NEST, Scuola Normale Superiore and Istituto di Nanoscienze-CNR, I-56126 Pisa, Italy}
\author{Martin Leib}
\affiliation{Technische Universit\"at M\"unchen, Physik Department, James-Franck-Str., D-85748 Garching, Germany}
\author{Michael J. Hartmann}
\affiliation{Technische Universit\"at M\"unchen, Physik Department, James-Franck-Str., D-85748 Garching, Germany}

\date{\today}

\begin{abstract}
  We introduce and study the properties of an array of QED cavities coupled by nonlinear elements, 
  in the presence of photon leakage and driven by a coherent source. 
  The nonlinear couplings lead to photon hopping and to nearest-neighbor Kerr terms.
  By tuning the system parameters, the steady state of the array can exhibit a {\it photon crystal} 
  associated to a periodic modulation of the photon blockade. In some cases the crystalline ordering 
  may coexist with phase synchronisation. The class of cavity arrays we consider can be built 
  with superconducting circuits of existing technology.
\end{abstract}



\maketitle

{\it Introduction.---} Since its beginning, the study of light-matter interaction in cavity 
and circuit QED has been providing a very fertile playground to test fundamental questions at the heart 
of quantum mechanics, together with the realisation of very promising implementations of 
quantum processors~\cite{raimond2001,girvin2009}. The coupling of separate cavities 
through photon hopping introduces an additional degree of freedom that is receiving increasing interest 
both theoretically and experimentally. 

Cavity arrays, periodic arrangements of neighbouring QED cavities, 
have been introduced~\cite{hartmann2006,greentree2006,angelakis2007} as prototype systems 
to study many-body states of light. Their very rich phenomenology arises from the interplay 
between strong local nonlinearities and photon hopping. 
In the photon blockade regime the array enters a Mott insulating phase, where photon number 
fluctuations are suppressed. In the opposite regime, where the hopping dominates, photons 
are delocalised through the whole array with long-range superfluid correlations. 
The phase diagram has been thoroughly studied by a variety of methods and the location 
of the different phases, together with the critical properties of the associated phase transitions, 
have been determined (see, {\it e.g.}, the reviews~\cite{hartmann2008,tomadin2010,houck2012}).

The properties of cavity arrays resemble in several aspects those of 
the Bose-Hubbard model~\cite{fisher1989}, as long as particle losses can be ignored. 
Cavity arrays, however, will naturally operate under nonequilibrium conditions, 
{\it i.e.} subject to unavoidable leakage of photons which are pumped back into the system 
by an external drive. In this case the situation may change drastically, and, 
to a large extent, it is an unexplored territory. 
Only very recently the many-body nonequilibrium dynamics of cavity arrays started to be 
addressed~\cite{carusotto2009,tomadin2010b,Hartmann10,nissen2012}, thus entering 
the exciting field of quantum phases and phase transitions in driven 
quantum open systems~\cite{diehl2008,diehl2010,baumann2010,lesanovsky2010,lee2011,ludwig2012}. 

Since the very beginning, all the works devoted to cavity arrays studied the case 
in which adjacent cavities are coupled by photon hopping. 
In this Letter we introduce a new class of arrays in which the coupling between cavities 
is mediated by a nonlinear element/medium.  Thanks to the flexibility in the design of the nonlinear 
coupling elements, these finite-range couplings can appear in the form of cross-Kerr nonlinearities 
and/or as a correlated photon hopping, leading to a steady-state phase diagram that is a lot 
richer than the cases which have been considered so far. Here we discuss in particular the appearance 
of a new phase in cavity arrays, a {\it photon crystal}, which emerges in the steady-state regime 
when the array is driven by a coherent homogeneous pump.

A technology that is very well suited for realizing cavity arrays with such features is provided by 
circuit QED~\cite{Leib10}, where exceptional light-matter coupling has been demonstrated~\cite{Niemczyk10}, 
first experiments with arrays of up to five cavities have been done~\cite{Lucero12}, 
and great progress towards experiments with lattices of cavities has already been achieved~\cite{houck2012}.

In the following we first introduce the model for the cavities with their nonlinear couplings, 
the external drive and the unavoidable leakage of photons.  We present a possible implementation 
in an array of circuit-QED cavities that are coupled via a nonlinear element.
We then study the steady-state regime by means of a mean-field approach and
Matrix Product Operator (MPO) simulations~\cite{HPCP09}. The scenario that emerges is rather complex, 
with the appearance of a number of phases and phase instabilities. 
We focus in particular on the possibility of spatial photon patterns that can emerge. 
For bipartite lattices the photon blockade is modulated on two different sublattices, 
furthermore on increasing the photon hopping it may also coexist with a global coherent state.

\begin{figure}
  \includegraphics[width = 1.0 \linewidth]{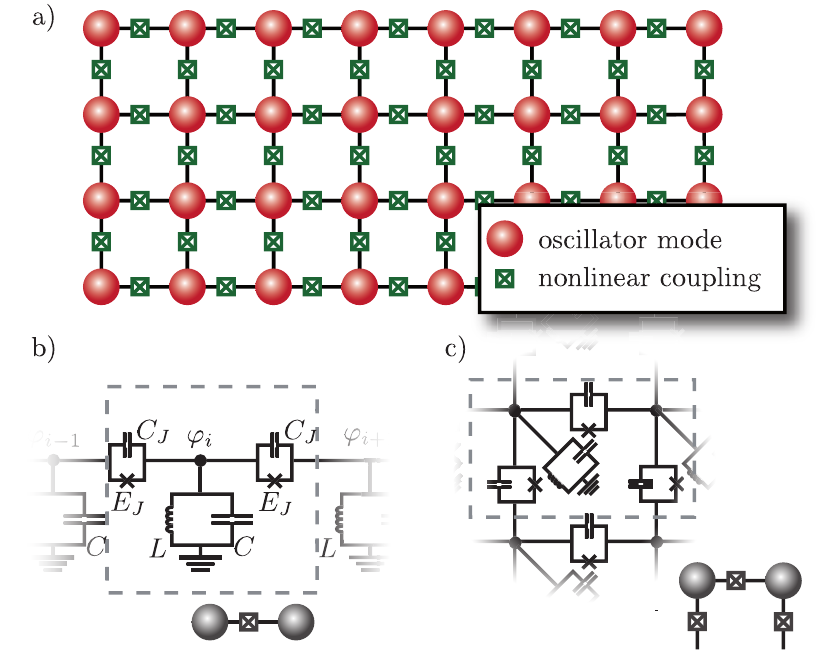}
  \caption{(color online). {\bf a}) An array of QED-cavities described by oscillator 
      modes (red circles) that are coupled via nonlinear elements (crossed boxes). 
      {\bf b}) and {\bf c}) Implementation of its building blocks in circuit-QED for one- and two-dimensional lattices.
      The circuit cavities are represented by a LC-circuit with capacitance $C$ 
      and inductance $L$ and mutually coupled through a Josephson nano-circuit, with 
      capacitance $C_{J}$ and Josephson energy $E_{J}$, that generates the on-site and cross-Kerr 
      terms in Eq.~(\ref{hamiltonian}). Details of this implementation can be found in Ref.~\cite{supplement}.
      An alternative approach to cross-Kerr interactions is discussed in Ref.~\cite{hu2011}. }
  \label{array}
\end{figure}

{\it The model.---} The cavity array is sketched in Fig.~\ref{array}a. The coupling between the cavities 
is mediated by a nonlinear element. In the specific implementation in circuit-QED this element 
is a Josephson nano-circuit.  When the coupling between the cavities is realized 
through the circuit described in Fig.~\ref{array}b, linear tunneling of photons between 
adjacent cavities can be tuned and even fully suppressed by adjusting the nonlinear coupling circuits 
to a suitable operating point, see Supplementary Material (SM)~\cite{supplement} for details. 
In this regime the cavities are coupled via a strong cross-Kerr term and further correlated-hopping terms,
which can lead to considerable modifications in the phase diagram~\cite{Sowitski2012}. 
Yet there are also more involved approaches involving multiple transmon qubits to realize 
cross-Kerr interactions in the absence of correlated hoppings~\cite{hu2011}, see also~\cite{zueco2012, peropadre2013}.
%
%
However, in the regime of parameters we are interested in and where a photon crystal emerges, correlated 
hopping leads to small quantitative corrections at the expense of complicating considerably the analysis. 
In the SM we quantify these differences in more details~\cite{supplement}. 
  
Here for simplicity, we concentrate on the salient features of the array in Fig.~\ref{array} 
which are captured by the effective Hamiltonian (in the rotating frame),
\begin{eqnarray}
  {\cal H} & = & \sum_i{[- \delta a_i^\dagger a_i+ \Omega(a_i^\dagger+a_i)]} -J \sum_{\langle i,j\rangle} (a_i^{\dagger} a_j+ 
  \mbox{H.c.})\nonumber \\
  & + & U \sum_i n_i(n_i-1) + V \sum_{\langle i,j\rangle} n_i n_j \; , 
  \label{hamiltonian}	
\end{eqnarray}
where the number operator $n_i =  a_i^\dagger a_i$ counts the photons in the $i$-th cavity 
($ a_i^\dagger/ a_i$ being the corresponding creation/annihilation operators).
The first three terms describe respectively the detuning $\delta$ of the cavity mode with respect 
to the frequency of the pump, the coherent pump with amplitude $\Omega$ and the hopping of photons 
between neighboring cavities at rate $J$.
The last two terms take into account 
the nonlinearities through the onsite- and cross-Kerr terms with the associated 
energy scales $U$ and $V$ respectively.  In the specific case of circuit-QED arrays, the two
types of nonlinearities can be realized through the setup of Fig.~\ref{array}. 
In order to keep our results as general as possible, we consider the effective model~(\ref{hamiltonian}) 
without specifying further the underlying matter-light interaction term. 

The dynamics of the array is governed by the Master equation
\begin{equation}
  \dot{\rho} = -i[{\cal H},\rho] + \frac{\kappa}{2} \sum_{i} (2 a_{i}\rho
  a_{i}^{\dag} - n_{i} \rho - \rho n_{i}) \;,
\label{me}
\end{equation}
where $\kappa ^{-1}$ is the photon lifetime in each cavity.
The model in Eq.~(\ref{hamiltonian}) together with Eq.~(\ref{me}) encompasses, in some limiting cases, 
regimes that were already addressed in the literature. The regime of $U \to \infty$ and $J=0$ 
was considered in Ref.~\cite{lee2011}, where an antiferromagnetic phase was first predicted in Rydberg atoms. 
The case of on-site Kerr nonlinearity, {\it i.e.} $V=0$, is the only one studied 
so far in cavity arrays~\cite{Hartmann10}. The model considered here offers a much richer phase diagram. 
A unique characteristics of the cavity arrays with nonlinear couplings is that 
the cross-Kerr nonlinearity $V$ can even exceed $U$. By coupling an additional qubit locally 
to each resonator~\cite{Neumeier12}, different ranges of the ratio $V/U$ can be explored. 
Moreover, in devices where on-chip control lines can be used to locally thread magnetic fields 
through the loops of the coupling circuits and the additional transmons, the ratios 
$J/U$ respectively $J/V$ and $V/U$ can be tuned on-chip. 
For this reason we will, in the following, consider $U$, $V$ and $J$ as independent.

We first discuss the steady-state phase diagram in the mean-field approximation, 
which becomes accurate in the limit of arrays with large coordination number $z$. 
The decoupling in Eq.~(\ref{hamiltonian}) is performed on the hopping and cross-Kerr terms,
$z^{-1} \sum_{\langle i,j\rangle} a_i^{\dagger} a_j \, \to \,
\langle a_A^{\dagger} \rangle \sum_{i \in B }  a_i + \langle a_B^{\dagger} \rangle \sum_{ j \in A  }  a_j$
and
$z^{-1}\sum_{\langle i,j\rangle} n_in_j \, \to \, 
\langle n_A \rangle \sum_{i \in B }  n_i + \langle n_B \rangle \sum_{ j \in A  }  n_j$,
where we assumed a bipartite lattice, $A$ and $B$ being the two sublattices. 
The mean-field analysis simplifies the dynamics dictated by Eq.~(\ref{me}) to two coupled 
equations for the two different sublattices.  As a function of all the parameters 
characterizing the system and its dynamics, one gets a very rich behavior 
in the asymptotic regime which includes steady-state/oscillating phases, as well as 
uniform/staggered configurations. Here we highlight what we think are its most intriguing features.
All the couplings will be expressed in units of the photon lifetime, $\kappa =1$.

\begin{figure}
  \includegraphics[width = 0.8\columnwidth]{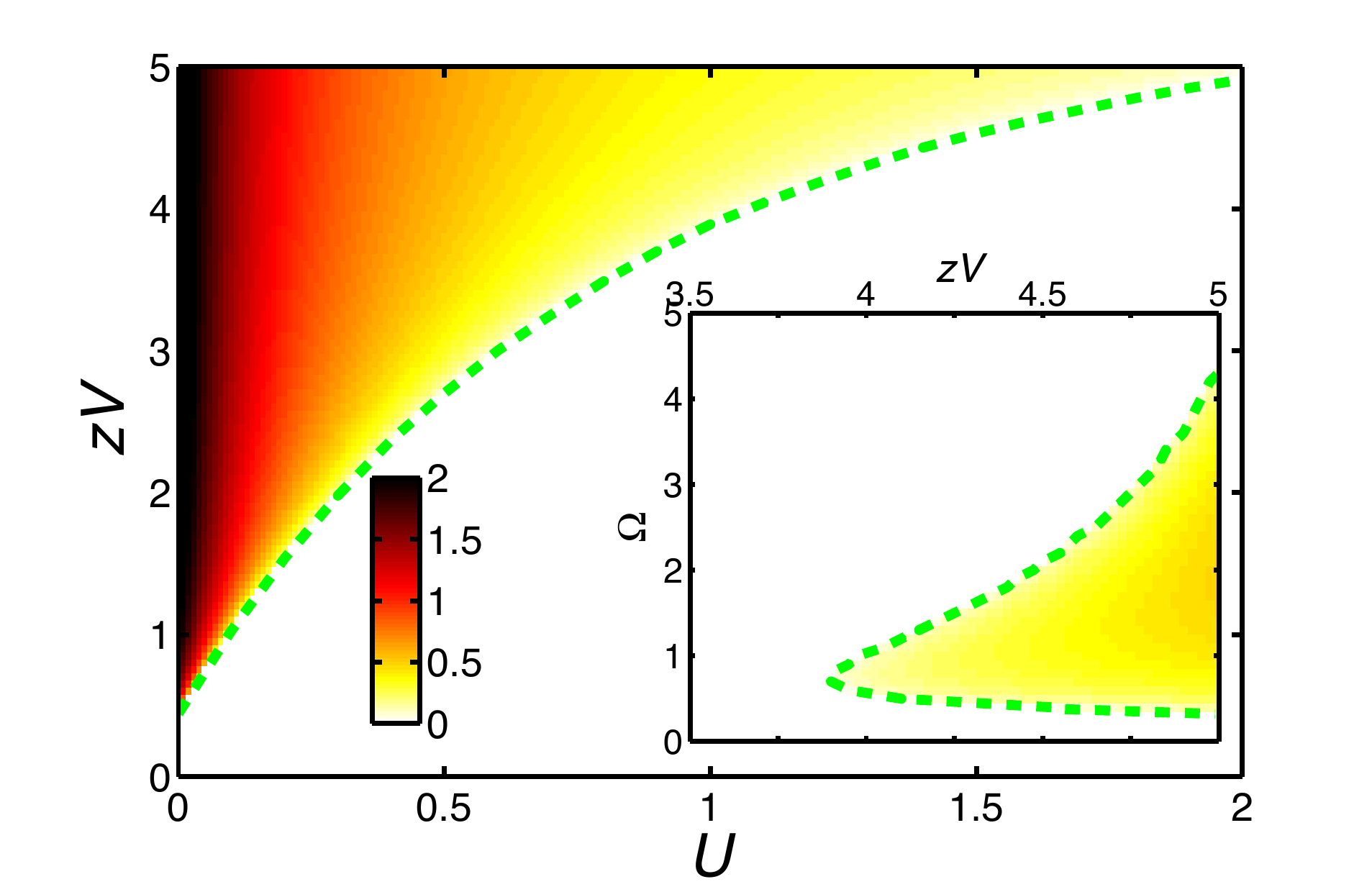}
  \caption{(color online). Order parameter $\Delta n$ for the photon crystal in the $U-V$ plane 
    at zero hopping. If the cross-Kerr term exceeds a critical threshold $V_c$, 
    the steady state is characterized by a staggered order in which $\Delta n \ne 0$. 
    Here we fixed $\Omega=0.75$ and $\delta = 0$, for which $z V_c \approx 0.44$ at $U=0$, while 
    $zV_c \approx 5.73$ in the hard-core limit ($U \to \infty$).
    In the inset we show $\Delta n$ as a function of $\Omega$ and $V$ at a fixed value of $U = 1$.
    Here and in the next figure the color code signals the intensity of the order parameter, 
    while dashed green lines are guides to the eye to locate the phase boundaries.}
  \label{crystalJ0}
\end{figure}

{\it Mean-field steady-state diagram.---} Fig.~\ref{crystalJ0}, where for the moment 
we set the hopping to zero, shows that, on increasing the cross-Kerr term, the array 
can reach a steady state in which the photon number is modulated as in a photon crystal, 
the order parameter being $\Delta n = \vert \langle n_A \rangle - \langle n_B \rangle \vert$. 
Here the area above the green line denotes the crystalline phase. 
In the $U \to \infty$ limit, the transition between the uniform and the crystal phase 
is located at $z V_c \simeq \gamma_\infty (-2 \delta + \sqrt{\gamma_\infty})/4\Omega^2$
with $\gamma_\infty = 4 \delta^2 + 8 \Omega^2 +1$, the previous expression holding for small detunings, 
and coincides with the transition to an antiferromagnetic phase described in Ref.~\cite{lee2011}.
Note that a lower value of $U$ favors the crystal phase. 
In the opposite limiting case of $U=0$, the transition is found 
at $z V_c\ \simeq \gamma_0 (-2 \delta + \sqrt{\gamma_0})/4\Omega^2$ with $\gamma_{0} = 4 \delta^2 + 1$.
We deliberately considered a regime in the parameter space where $V \ge U$ since, 
as already mentioned, it is a peculiar feature of the cavity arrays proposed here.
The transition to the crystal phase is reentrant as a function of the drive (inset to Fig.~\ref{crystalJ0}). 
At very small pumping the density is too low to lead to a photon crystal. 
Vice-versa it also disappears on increasing $\Omega$, since pumping favors an homogeneous photon arrangement.
A similar feature has been observed in the limit $U = \infty$~\cite{lee2011}. 

\begin{figure}
  \includegraphics[width=0.79 \columnwidth]{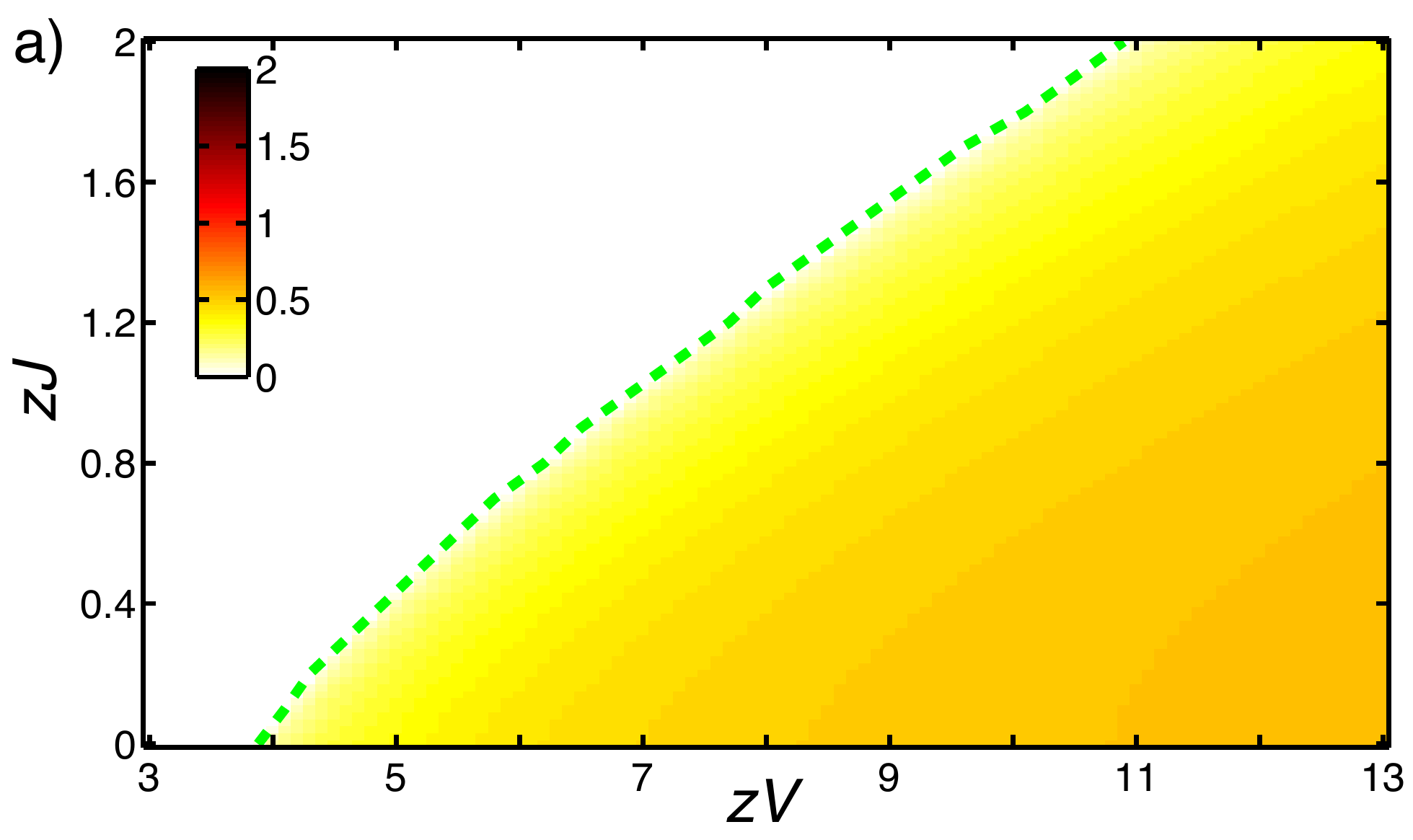}
  \includegraphics[width=0.8 \columnwidth]{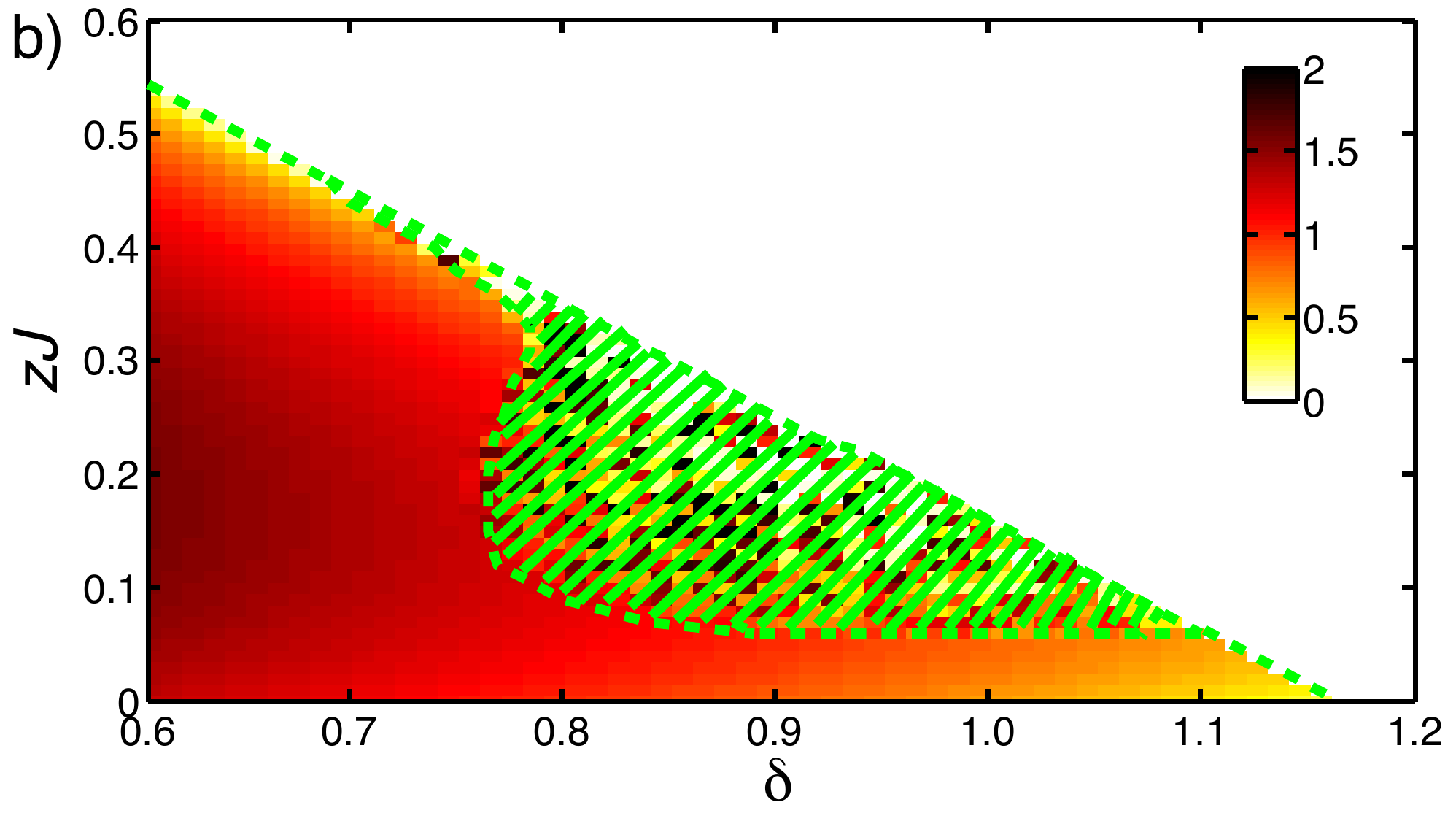}
  \caption{(color online). Order parameter for the photon crystal at finite hopping.
    $\Delta n$ is plotted as a function of $J$ and $V$, for $\delta=0$ and $U = 1$ (panel a), and
    as a function of $J$ and $\delta$, for $zV = 0.6$ and $U = 0$ (panel b). 
    Here we fixed $\Omega = 0.75$.  At finite values of the detuning, 
    in addition to the normal (white) and crystalline (colored) phases, an intermediate region
    (shaded green), characterized by an oscillatory behavior in the asymptotic state, appears.
    As discussed in the main text, we suggest this last regime may be seen 
    as a nonequilibrium analog of a supersolid.}
  \label{crystaldelta}
\end{figure}

If the hopping between photons is switched on, delocalization will suppress the solid phase 
and at a critical value of $J$ (which depends on $V,U, \delta$, and $\Omega$) 
there is a transition to a normal phase. This is shown in Fig.~\ref{crystaldelta}, 
as a function of the cross-Kerr nonlinearity $V$ (Fig.~\ref{crystaldelta}a) 
and of the detuning $\delta$ (Fig.~\ref{crystaldelta}b).
In Fig.~\ref{crystaldelta}a we display the case $U=1$, while at smaller values of $U$
the phase diagram shows a reentrance. Although interesting, further analysis is needed 
to see if this feature is only present at mean-field level. Yet it does not seem 
improbable that an increased hopping could facilitate the redistribution of particles 
into a crystalline order imposed by the interactions.
We conclude this discussion by pointing out that, as discussed in Ref.~\cite{supplement},
under nonequilibrium conditions it is even possible, although much harder, 
to realize a crystalline phase at $V=0$. As a matter of fact in that case, for some values 
of the coupling constants, the steady state can be either uniform or crystalline, 
depending on the initial conditions.

\begin{figure}
  \includegraphics[width=1. \columnwidth]{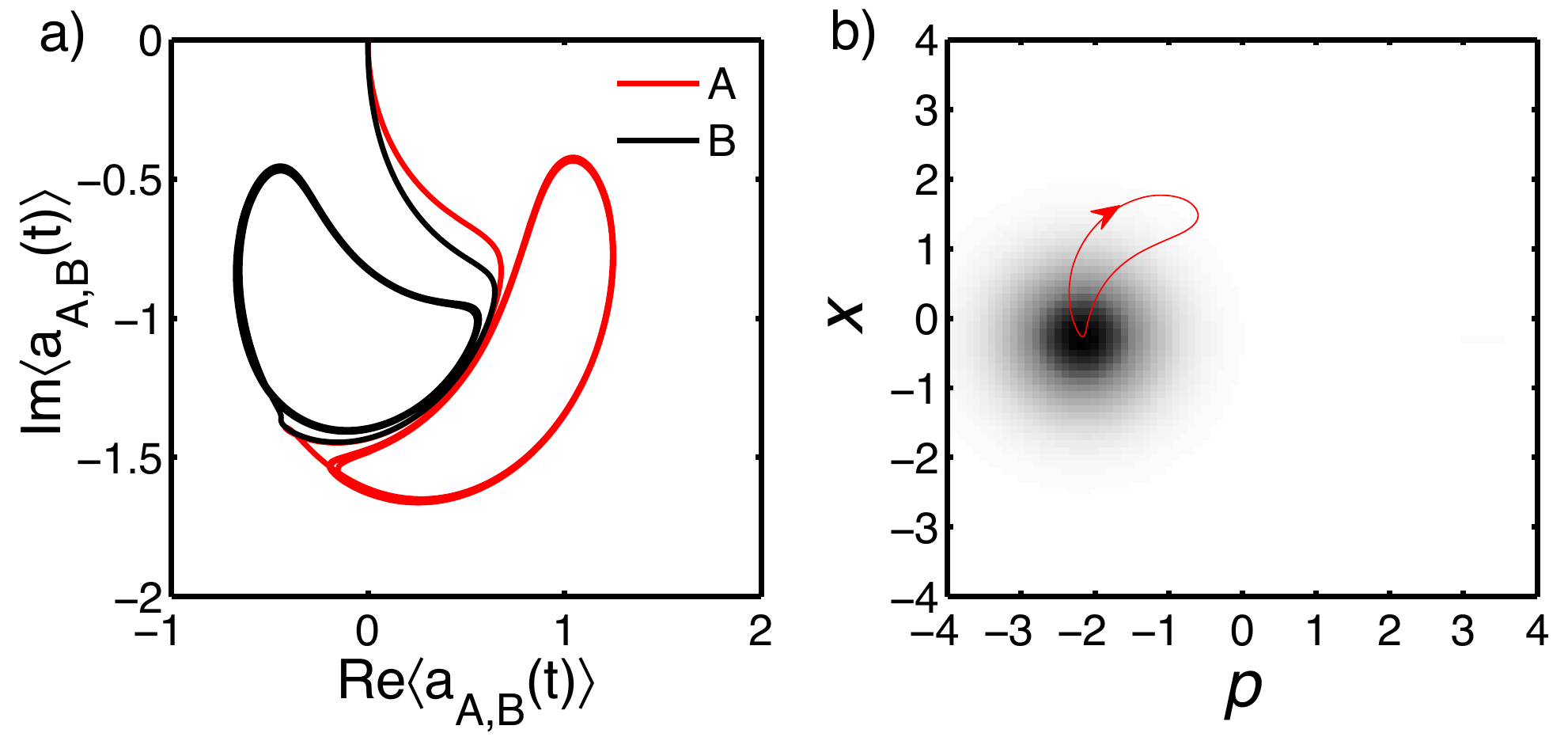}
  \caption{(color online). {\bf a}) Time dependent traces of the real and imaginary parts 
    of $\langle a \rangle$ for the two sublattices as a function of time in the steady-state regime 
    of the intermediate oscillating phase of Fig.~\ref{crystaldelta}b.
    {\bf b}) The Wigner transform of the reduced single-site density matrix for sublattice A
    is plotted at a given time in the intermediate oscillating phase of Fig.~\ref{crystaldelta}b.
    Here we used the same parameters as in Fig.~\ref{crystaldelta}b and fixed $zJ = 0.2$, $\delta = 0.9$.}
  \label{oscillating}
\end{figure}

The mean-field phase diagram in the $\delta-J$ plane and for $U=0$ is depicted in Fig.~\ref{crystaldelta}b. 
As highlighted in the region between the dashed green lines, for $0.8 \lesssim \delta \lesssim 1.1$, 
on switching on the photon hopping, a new intermediate phase appears.
In this region, even in the long-time limit the state never becomes completely stationary and
there is a residual time-dependence of $\langle a\rangle$ 
with $\langle a_A\rangle \ne \langle a_B\rangle$, {\it i.e.} there is an additional time dependence 
of $\langle a\rangle$ on top of the trivial oscillation with the frequency of the coherent drive 
that is hidden in our choice of the rotating frame. At the same time the system shows $\Delta n \ne 0$.
In Fig.~\ref{oscillating}a we show 
the time evolution of the real and imaginary parts of $\langle a\rangle$ for the two sublattices. 
A closer inspection of the properties of the oscillating phase reveals that the reduced 
density matrix of a single site (in either of the two sublattices) is a coherent state 
which evolves periodically in time as shown in Fig.~\ref{oscillating}b. There we plotted 
the Wigner function $W(x,p) = \int \langle x-y \vert \rho_{A} \vert x+y \rangle e^{2ipy} {\rm d} y$
of one sublattice at a given time, with $x = (a + a^\dagger)/\sqrt{2}$, $p= i(a^\dagger - a)/\sqrt{2}$, 
and $\vert x \rangle$ being an eigenstate of the position operator $x$. 
Following the analysis performed in Ref.~\cite{ludwig2012}, we are lead to conclude that in this region 
the dynamical evolution of the whole array is synchronised separately in the two different sublattices. 
The contemporary presence of checkerboard ordering and global dynamical phase coherence suggests us 
to view this intermediate phase as a {\it nonequilibrium supersolid phase}~\cite{boninsegni2012}.
The intermediate region extends also at finite-$U$ values, although the coherent state 
of Fig.~\ref{oscillating}b will be progressively deformed on increasing the on-site repulsion.

\begin{figure}
  \includegraphics[width=0.9\columnwidth]{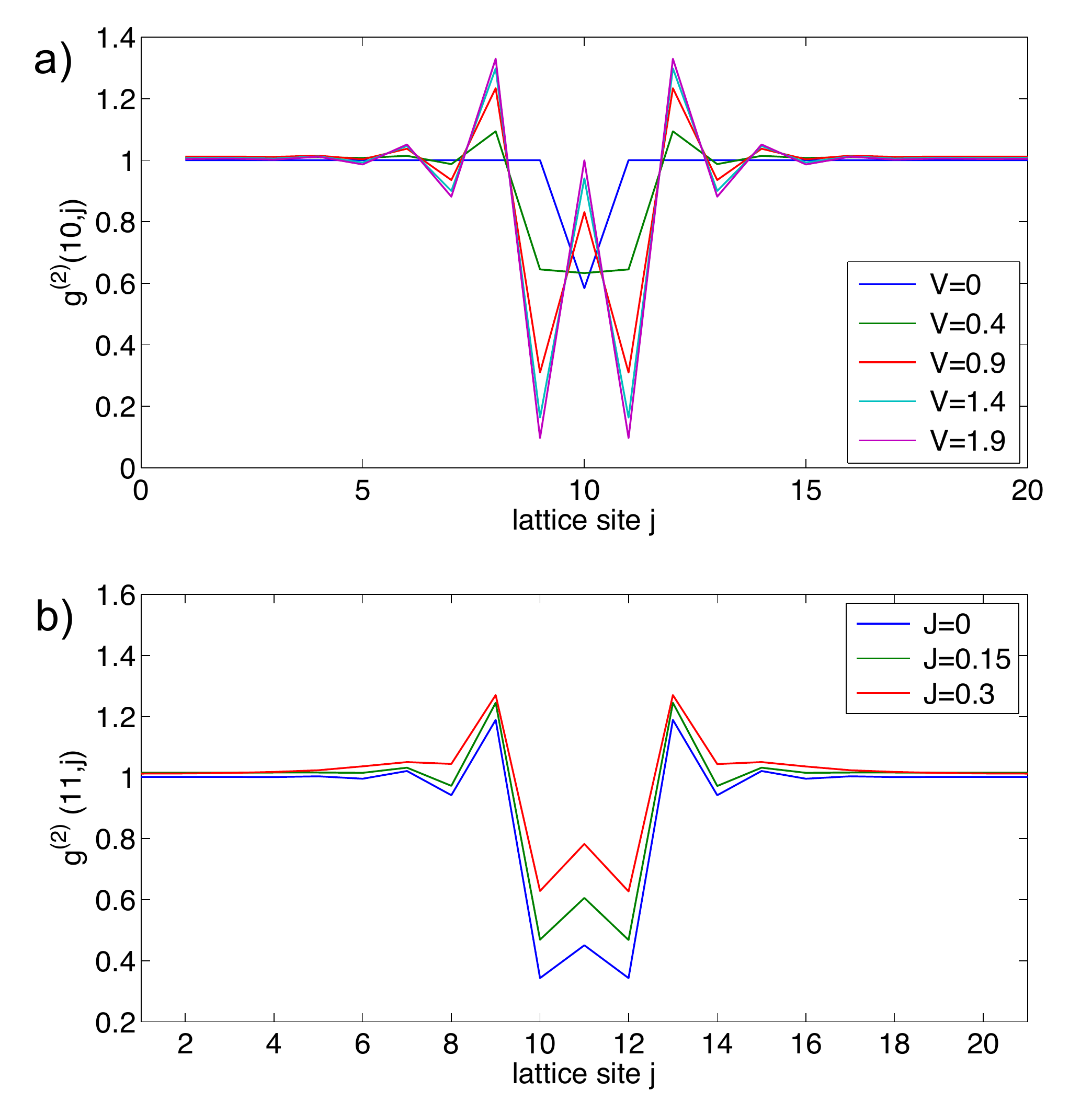}
  \caption{(color online). MPO results for linear chains of cavities. 
    {\bf a}) $g^{(2)}(10,j)$ for $\delta = 0$, $J = 0$, $U = 0.5$, $\Omega = 0.4$ and $V$ as in the legend.
    {\bf b}) $g^{(2)}(11,j)$ for $\delta = 0$, $U = 1$, $V = 1$, $\Omega = 0.4$ and $J$ as in the legend.}
  \label{DMRG}
\end{figure}

{\it MPO simulations.---} Most of the features we discussed can already 
be seen in small arrays. To show some examples and to further support the mean-field analysis 
given above, we here present results that were obtained for linear chains of cavities, $z=2$, with 
MPO simulations~\cite{Hartmann10} of the Master equation~(\ref{me}), which provide a (numerically) exact 
description of its nonequilibrium many-body dynamics. 
Fig.~\ref{DMRG}a shows the density-density correlation function 
$g^{(2)}(i,j) = \langle a_{i}^{\dagger} a_{j}^{\dagger} a_{j} a_{i} \rangle / \langle n_{i} \rangle \langle n_{j}\rangle$ 
for a chain of 20 cavities with $\delta = 0$, $J = 0$, $U = 0.5$, $\Omega = 0.4$
and various values of $V$. One clearly sees that a staggered dependence of the distance $r=|i-j|$, 
indicating strong density-density correlations, appears for nonzero $V$, whereas for $V=0$ photons 
in distinct cavities are uncorrelated ($g^{(2)} = 1$). Fig.~\ref{DMRG}b shows $g^{(2)}$ for a chain 
of 21 cavities with $\delta = 0$, $U = 1$, $V = 1$, $\Omega = 0.4$ and various values of $J$. 
The spatial range of density-density correlations shrinks with increasing tunneling rate $J$, 
indicating a crossover to an uncorrelated state. A more quantitative analysis of the decay
of correlations with increasing distance is not conclusive for the chain length considered here. 
A true ordering in the steady state can probably only be stabilized in two-dimensions.

{\it Conclusions.---} In this Letter we introduced cavity arrays with coupling mediated by nonlinear elements. 
This opens the way to study a variety of new possibilities, including correlated photon hopping 
and finite-range photon blockade.  We concentrated on this last point studying the effect 
of a cross-Kerr nonlinearity on the steady state and found a very rich phase diagram. 
A photon solid characterized by a checkerboard ordering of the average photon number appears 
for a substantial range of the coupling constants.
In addition we see that, for some choice of the parameters, a finite hopping stabilizes 
a phase where the crystalline ordering coexists with a globally synchronized dynamics of the cavities, 
suggesting an analogy to a nonequilibrium supersolid. 
Most of the results presented in this work were obtained in a mean-field approximation. 
We corroborated the existence of a steady-state solid phase by studying a one-dimensional array 
by means of a Matrix Product Operator approach. This last analysis confirms that a
crystalline ordering of photons can be observed with existing experimental technology. 

{\it Acknowledgments.---} We acknowledge fruitful discussions with A. Tomadin. 
This paper was supported by EU - through Grant Agreement No. 234970-NANOCTM, and No. 248629-SOLID, 
by DFG - through the Emmy Noether project HA 5593/1-1 and the CRC 631 and by 
National Natural Science Foundation of China under Grant No. 11175033.

\widetext

\appendix

\section{Supplementary Material} 

\subsection{Cross-Kerr nonlinearities in arrays of circuit cavities}

Here we describe a circuit quantum electrodynamics (cicuit-QED) setup that represents an array 
of cavities coupled by nonlinear elements and can be effectively described by a lattice of harmonic oscillators 
coupled via cross-Kerr nonlinearities. The setup thus provides an experimentally feasible way 
for implementing the model described by equations (1) and (2) of the main text. 
Cicuit-QED is particularly well suited for this task because of the great design flexibility, 
the tunable nonlinearity provided by Josephson junctions and the exceptionally 
high coupling between subsystems that can be reached.

\begin{figure}[b]
  \includegraphics{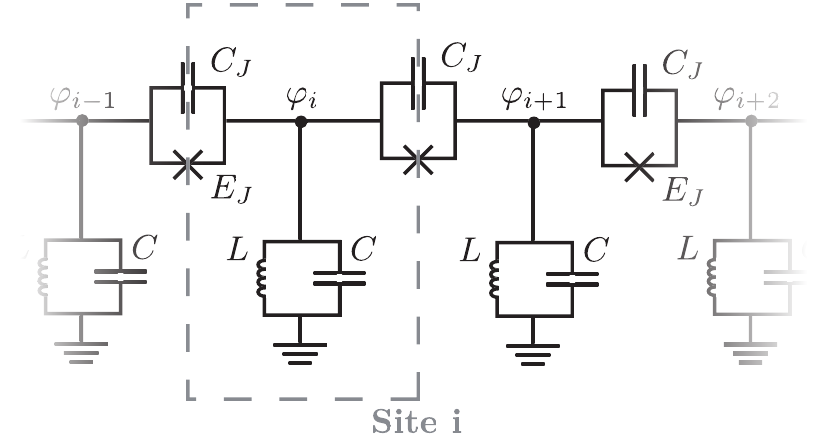}
  \caption{(color online). Electrical circuit sketch of the setup we envision to realize a system 
    with cross Kerr nonlinearities. Here for the example of a one-dimensional chain of cavities. 
    Lumped element LC-circuits are coupled via capacitively shunted Josephson junctions. 
    The Josephson energies could be tuned {\it in situ} by replacing the Josephson junctions 
    with superconducting interference devices.}
  \label{crossKerrSetup}
\end{figure}

For our derivation, we consider lumped element resonators, representing the cavities 
in our array, which are coupled conductively through capacitively shunted Josephson junctions, 
see Fig.~\ref{crossKerrSetup} for a sketch of the circuit representing a linear chain of cavities.
We focus on lumped element resonators, in order to keep the derivation simple and transparent. 
Coplanar waveguide resonators work equally well. For the considered combination of capacitive 
and inductive coupling between the resonators, the linear parts of the couplings can cancel 
each other for suitable choices of the parameters, leaving a residual coupling 
via the nonlinearity of the Josephson junctions. 

In terms of the node fluxes $\phi_i$, the Lagrangian of the whole setup reads,
\begin{equation}
  \mathcal{L}=\sum_i\left[\frac{C}{2}\dot{\phi}_i^2-\frac{1}{2 L}\phi_i^2\right]
  +\sum\limits_{\langle i,j\rangle}\left[\frac{C_J}{2}(\dot{\phi}_i-\dot{\phi}_{j})^2+E_J\cos\left(\frac{\phi_i-\phi_{j}}{\phi_0}\right)\right], 
\end{equation}
with $L$ and $C$ the inductance and capacitance of the lumped element resonators and $C_J$ and $E_J$ the capacitance and Josephson energy of the Josephson junctions.
$\phi_0 = \hbar/(2 e)$ is the reduced flux quantum.
Assuming $C_J/C\ll1$ we invert the capacitance matrix to first order in $C_J/C$ for performing the Legendre transformation to obtain the Hamiltonian~\cite{Nunnenkamp2011},
\begin{eqnarray}
  \mathcal{H}&=&\sum\limits_i\left(\frac{q_i^2}{2\tilde{C}}+\frac{\phi_i^2}{2 \tilde{L}}\right)-\mathcal{H}_{c}-\mathcal{H}_{nl}\\
  \mathcal{H}_{c}&=&\sum\limits_{\langle i,j\rangle} \left(\frac{1}{\tilde{C}}\frac{C_J}{C+2C_J}q_iq_{j}-\frac{1}{\tilde{L}}\frac{2 L}{2 L + L_J}\phi_i\phi_{j}\right)\nonumber\\
  \mathcal{H}_{nl}&=&E_J\sum\limits_{\langle i,j\rangle}\left(\sum\limits_{n=2}^{\infty}(-1)^n\frac{(\phi_i-\phi_{j})^{2n}}{(2n)!\,\phi_0^{2n}}\right), \nonumber
\end{eqnarray}
where $\tilde{C}=C+2C_J$, $1/\tilde{L}=1/(2L)+1/L_J$, and $L_{J} = \phi_{0}^{2}/E_{J}$.

The charges on the islands $q_i = C \dot{\phi}_{i} - C_{J}\sum_{j \langle i \rangle}(\dot{\phi}_{i}+\dot{\phi}_{j \langle i \rangle})$ 
($j \langle i \rangle$ denotes all connections to site $i$), 
defined by the coupling capacitances of the Josephson junctions and the lumped element resonators, 
and the fluxes $\phi_{i}$ associated to the phase drop at the inductance of the lumped element 
resonators are our canonically conjugate variables. We quantize the Hamiltonian by introducing bosonic 
lowering and raising operators $a_i$ and $a_i^{\dag}$ according to,
\begin{eqnarray}
  \phi_i&=&\frac{1}{\sqrt{2}}\left(\frac{\tilde{L}}{\tilde{C}}\right)^{\frac{1}{4}}(a_i+a_i^{\dag})\\
  q_i&=&\frac{i}{\sqrt{2}}\left(\frac{\tilde{C}}{\tilde{L}}\right)^{\frac{1}{4}}(a_i^{\dag}-a_i)\,.
\end{eqnarray}
The quantized coupling Hamilton operator thus reads,
\begin{equation}
  H_{c}=\sum\limits_{\langle i,j\rangle}\omega X_{-}(a_i^{\dag}a_{j}+a_ia_{j}^{\dag})+\sum\limits_{\langle i,j\rangle}\omega X_{+} (a_i^{\dag}a_{j}^{\dag}+a_ia_{j}), \nonumber
\end{equation}
where the oscillator frequency is $\omega=1/\sqrt{\tilde{L}\tilde{C}}$ and 
$X_{\pm} = \frac{2 L}{2 L + L_J} \pm \frac{C_J}{C+2C_J}$. We choose $C_J/(C+2C_J)=2L/(2L+L_J)=\alpha$ 
such that $X_{-} = 0$ and the linear tunneling between neighboring oscillator sites vanishes. 
Furthermore we neglect the terms proportional to $X_{+}$ in a rotating wave approximation 
which is justified as long as $C_J/(C+2C_J)+2L/(2L+L_J)\ll2$ holds. 
In this way we arrive at a chain of harmonic oscillators that are decoupled in linear order.

The nonlinear parts of the Hamilton operator however provide us with coupling terms between 
the neighboring oscillators and on-site nonlinearities.  
We restrict ourselves to fourth order nonlinearities, perform a rotating wave approximation 
and arrive at the many-body Hamiltonian,
\begin{equation} \label{eq:Heff}
  H = \sum_{i} \left[(\omega+\delta\omega) a_i^{\dag}a_i -\alpha z \frac{E_C}{2} a_i^{\dag}a_i^{\dag}a_i a_i\right] + \sum_{\langle i,j \rangle} \left[-2 \alpha E_C a_i^{\dag}a_ia_{j}^{\dag}a_{j} + H_{i,j}^{(ch)}\right],
\end{equation}
where $\delta\omega$ is a small correction coming from the normal ordering process 
of the nonlinearity and $E_C=e^2/(2 \tilde{C}^2)$ is the charging energy of the individual 
lattice sites of our model. The second term in Eq.~(\ref{eq:Heff}) is an on-site Kerr term 
whereas the third term describes cross-Kerr interactions.
This specific implementation thus realizes the Hamiltonian of equation (1) in the main text 
for $U = - \alpha z E_{C}/2$ and $V = - 2 \alpha E_{C}$, thus leading to $z V/U = 4$. 

\begin{figure}[h]
  \includegraphics[width=0.6 \columnwidth]{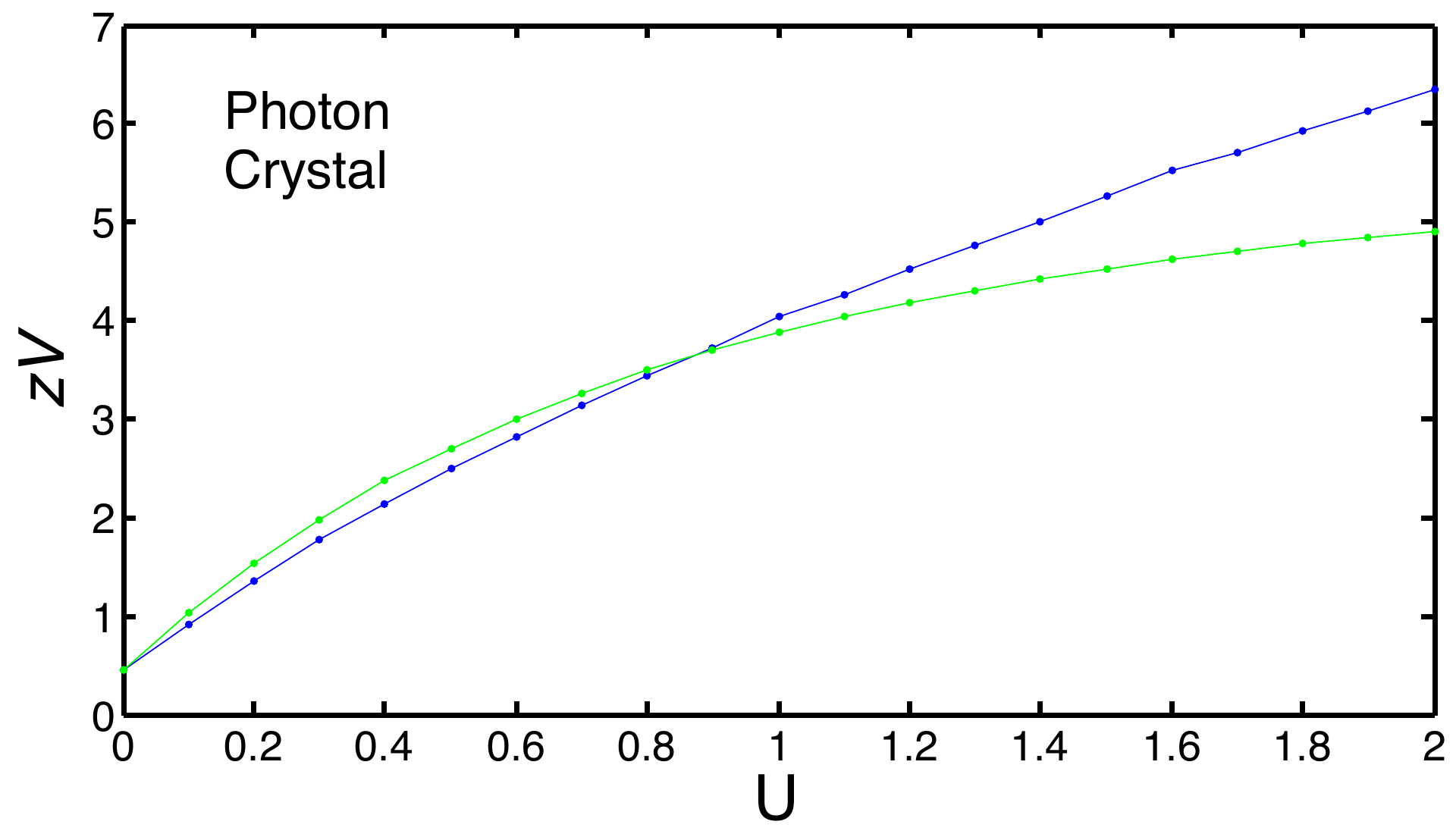}
  \caption{(color online). Emergence of the photon crystal in the $U-V$ diagram at $J=0$.
    We highlight the effects induced by the terms in $H_{i,i+1}^{(ch)}$ by comparing mean-field results 
    without (green line) and with (blue line) such contributions.
    Here the strength of the correlated-hopping terms is set equal to $-U$,
    while we set $\Omega = 0.75$ and $\delta=0$, as in Fig.~\ref{crystalJ0} of the main text.}
  \label{corr_hop}
\end{figure}

The term $H_{i,j}^{(ch)} = \alpha E_C\big(a_ia_{j}^{\dag}a_{j}^{\dag}a_{j} + a_i^{\dag}a_i^{\dag}a_ia_{j} 
-\frac{1}{2}a_i^{\dag}a_i^{\dag}a_{j}a_{j}\big) + \text{H.c.}$ describes correlated hopping 
of photons between adjacent sites $i$ and $j$. For the purposes of the present work, {\it i.e.} a description 
of solid and supersolid phases, this term can be treated at a mean-field level.
A mean-field approximation would replace $a_i^{\dag}a_i \to \langle a_i^{\dag}a_i \rangle$ 
and the first two terms will thus only give rise to a (possibly sublattice dependent) renormalization 
of the single-photon hopping. The term $a_i^{\dag}a_i^{\dag}a_{i+1}a_{i+1}$ in turn is a factor of 4 weaker 
than the cross-Kerr term. Moreover, in regimes of large Kerr interactions, double occupations 
of lattice sites are small and this term becomes ineffective. For these reasons, in most of our
analysis we disregard the correlated hopping at present. 
A signature of the effects induced by $H_{i,i+1}^{(ch)}$ is evident from the data 
shown in Fig.~\ref{corr_hop}, where we plotted the phase boundary between
normal and crystalline phase in the $U-V$ plane at zero hopping.
The relatively small discrepancies in the curves with and without such terms points towards 
the fact that only quantitative modifications are induced in the parameter space we are considering, 
while all the qualitative features of our results should be unaffected.
It should however be noted that a more accurate treatment may lead to new phases in the diagram 
such as a pair superfluid state, when the corresponding hopping term becomes comparable 
with the cross-Kerr nonlinearity.
This is however beyond the scope of the present analysis and it will be treated separately. 

Our approximations require that $\alpha \ll 1$. Nonetheless the cross-Kerr interaction 
can be much larger than photon losses, {\it i.e.} $2 \alpha E_C \gg \kappa$, since, {\it e.g.}, transmon qubits 
have $E_{C}/h \sim 0.5$GHz and $T_{1} \sim 1\mu$s~\cite{Fedorov12}. 

Finally let us point out that the Josephson junctions linking two neighboring oscillators can be built tunable 
by replacing them with a dc-SQUID. In this way the $E_{J}$ and thus the $L_{J}$ can be modulated 
by applying an external flux to the dc-SQUIDs and the Hamiltonian~(\ref{eq:Heff}) can be tuned in real time. 
Hence, by choosing the external flux such that $X_{-} \not=0$ a linear tunneling of photons 
between the resonators can be introduced. Moreover larger on-site nonlinearities can be introduced 
by coupling each resonator locally to a superconducting qubit, {\it e.g}. a transmon.

\subsection{Photon crystal at $V=0$}

\begin{figure}
  \includegraphics[width=0.4 \columnwidth]{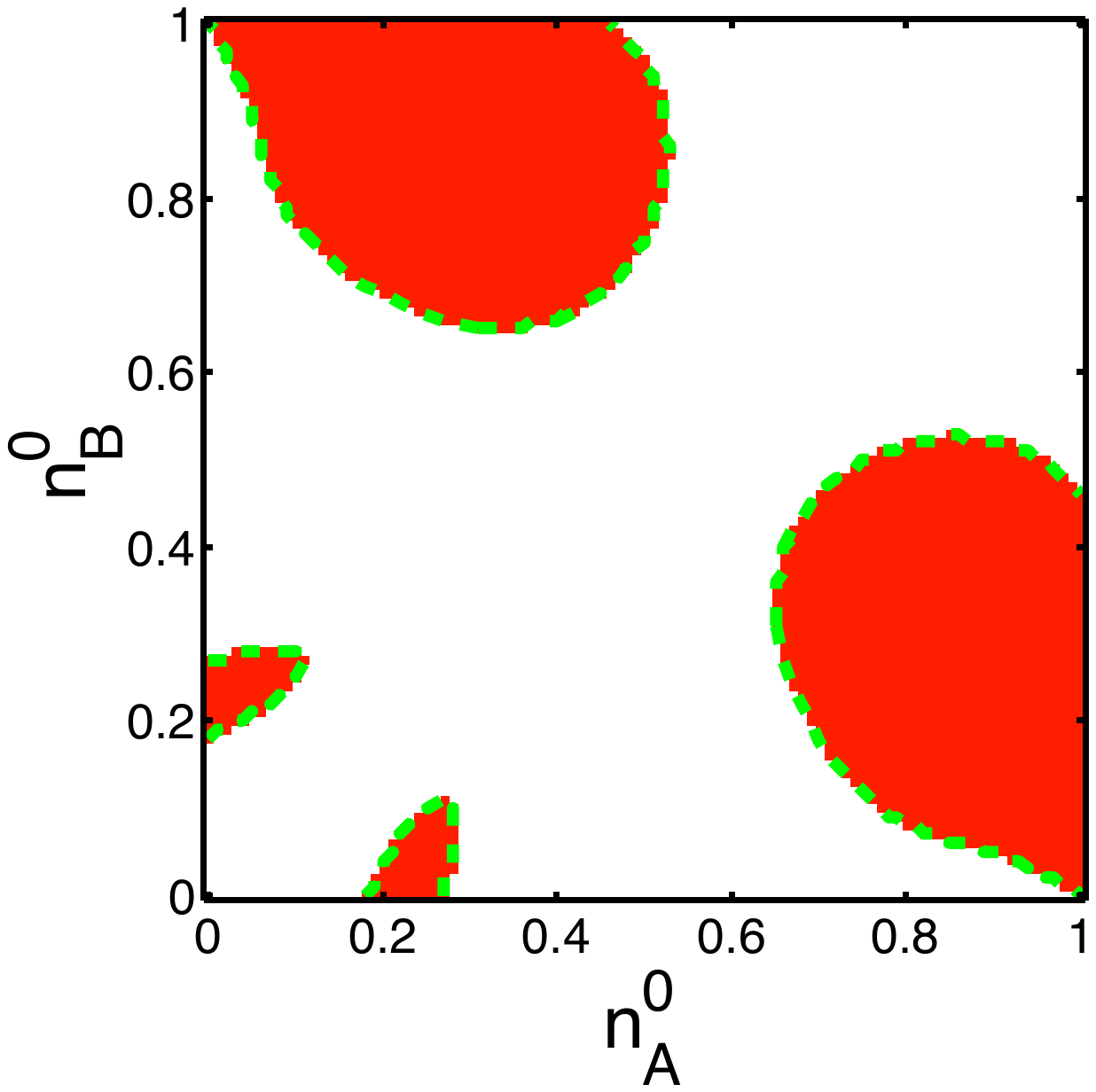}
  \caption{(color online). Phase diagram at $V=0$ in the space of initial conditions, with 
    $zJ = 6.2$, $\delta = 0.8$, $\Omega = 2$ and in the hard-core limit $U \to \infty$. 
    White areas denote starting occupation values where a steady-state uniform phase is reached, 
    while red areas denote initial conditions leading to an antiferromagnetic phase
    with $\Delta n \approx 0.1027$.}
  \label{crystalV0}
\end{figure}

We conclude the discussion of the solid phase by showing that the realization of a steady-state 
photon solid does not necessarily requires a cross-Kerr nonlinearity. 
Surely the presence of $V \ne 0$ stabilizes the solid in a wide range of the parameter space, thus 
making it easier to be observed experimentally. Nevertheless, while in equilibrium $V \ne 0$ 
is a necessary requirement, under nonequilibrium condition this ceases to be the case. 
The only requirement is a finite photon hopping and an initial unbalance in the occupation of the 
two sublattices.  
In Fig.~\ref{crystalV0} the regions in which the steady-state solid phase is reached are plotted 
as a function of the initial unbalance. Note that the value of the crystalline order parameter
$\Delta n$ does not depend on the choice of the initial conditions.

\end{document}